\documentclass[a4paper,onecolumn,11pt,accepted=2025-05-20]{quantumarticle}
\pdfoutput=1
\usepackage[utf8]{inputenc}
\usepackage[english]{babel}
\usepackage[T1]{fontenc}
\usepackage{amsmath}
\usepackage{amssymb}
\usepackage{amsfonts}
\usepackage{amsthm}
\usepackage{bm}
\usepackage[numbers,sort&compress]{natbib}
\usepackage{hyperref}
\hypersetup{
    colorlinks=true,
    linkcolor=blue,
    filecolor=blue,      
    urlcolor=blue,
    citecolor=blue,
    pdftitle={The State Preparation of Multivariate Normal Distributions using Tree Tensor Network},
    pdfpagemode=FullScreen,
}
\usepackage{braket}

\usepackage{tikz}
\usepackage{lipsum}

\DeclareMathOperator{\rank}{rank}
\newtheorem{theorem}{Theorem}[section]
\theoremstyle{definition}  % Examples are often styled like definitions
\newtheorem{example}{Example}[section]  % Numbered within each section

\newcommand{\norm}[1]{\left\lVert#1\right\rVert}

\begin{document}

\title{The State Preparation of Multivariate Normal Distributions using Tree Tensor Network}

\author{Hidetaka Manabe}
\affiliation{Graduate School of Engineering Science, Osaka University}
\orcid{0009-0002-2302-0357}
\author{Yuichi Sano}
\affiliation{Department of Nuclear Engineering, Kyoto University}
\maketitle

\begin{abstract}
The quantum state preparation of probability distributions is an important subroutine for many quantum algorithms. 
When embedding $D$-dimensional multivariate probability distributions by discretizing each dimension into $2^n$ points, we need a state preparation circuit comprising a total of $nD$ qubits, which is often difficult to compile.
In this study, we propose a scalable method to generate state preparation circuits for $D$-dimensional multivariate normal distributions,
utilizing tree tensor networks (TTN).
%We represent the probability distribution with a TTN and perform the task of quantum circuit compilation through the optimization of tensor networks.
We establish theoretical guarantees that multivariate normal distributions with 1D correlation structures can be efficiently represented using TTN. 
Based on these analyses, we propose a compilation method that uses automatic structural optimization to find the most efficient network structure and compact circuit.
We apply our method to state preparation circuits for various high-dimensional random multivariate normal distributions.
The numerical results suggest that our method can dramatically reduce the circuit depth and CNOT count while maintaining fidelity compared to existing approaches.

%Especially, by employing structural optimization, we can search for a network structure that efficiently represents the correlations between variables.
%The numerical results suggest that our method can dramatically reduce the circuit depth while maintaining fidelity compared to existing approaches.
%Moreover, for normal distributions with one-dimensional correlations, we can construct state preparation circuits in a scalable manner, regardless of the number of variables, by using tensor cross interpolation.
\end{abstract}

\maketitle

\section{Introduction}
Amplitude embedding of classical data is a key subroutine in various quantum algorithms such as quantum-linear-solver algorithms~\cite{harrow_Quantum_2009}, Monte-Carlo methods~\cite{montanaro_Quantum_2015}, 
and quantum machine learning~\cite{biamonte_Quantum_2017,mitarai_Quantum_2018}.
In quantum state preparation, we encode $2^n$ classical data $f(x)$ into an $n$-qubit quantum state $\ket{f(x)}=U(x)\ket0$.
Generally, achieving this requires $O(2^n)$ quantum operations~\cite{mottonen_Transformation_2004, plesch_Quantumstate_2011}, which can become a critical bottleneck in the execution of algorithms.
Thus, proposing efficient state preparation methods is essential for the practical implementation of quantum computer algorithms.
%Especially, the preparation of probability distributions is widely employed in various settings such as the Monte-Calo algorithm, finance, physics and so on~\cite{jordan_Quantum_2012,chowdhury_Quantum_2017,egger_Credit_2021,wilkens_Quantum_2023}. 

%In the realm of NISQ (Noisy Intermediate-Scale Quantum) devices, applications such as [specific examples] are notable. 
%In the context of Fault-Tolerant Quantum Computing (FTQC), examples include [specific examples].

There are many studies on methods of quantum state preparation~\cite{kitaev_Wavefunction_2009, carreravazquez_Efficient_2021, bauer_Practical_2021, rattew_Efficient_2021, rattew_Preparing_2022, nakaji_Approximate_2022, mcardle_Quantum_2022, zhu_Generative_2022, marin-sanchez_Quantum_2023, moosa_Lineardepth_2023, zylberman_Efficient_2023}. 
For the Fault-Tolerant Quantum Computer (FTQC), it is possible to utilize Quantum Signal Processing-based~\cite{mcardle_Quantum_2022} or
oracle-based algorithms, where error bounds and the scaling of the gate counts can be rigorously proven.
However, in the case of  NISQ (Noisy Intermediate-Scale Quantum devices) and early-FTQC, 
there are strict limitations on executable circuit depth, connectivity, and the number of qubits.
Therefore, it becomes essential to use classical optimizations 
%for specific algorithms and architectures
to drastically reduce the circuit costs.

One of the efficient compilation methods for state preparation circuits is the Fourier series loader (FSL) method~\cite{moosa_Lineardepth_2023}.
Suppose we want to encode a $D$-dimensional function $f(\bm{x})$, discretizing each dimension into $2^n$ points.
Implementing this would require a state preparation circuit with a $nD$-qubit circuits and a depth of $O(2^{nD})$. 
The FSL method utilizes the exponential decay of the Fourier coefficients of smooth functions, allowing for state preparation in 
a depth of $O(2^{mD})$ with $m<n$ using Fourier series approximation. 
Nevertheless, as $D$ increases, the circuit depth increases exponentially, making this method unsuitable for high-dimensional functions.

%In this study, we propose a tensor network-based method to compile the state preparation circuit of multivariate normal distributions,
In this study, we propose a new method to compile the state preparation circuit for multivariate functions, particularly for multivariate normal distributions,
which are widely used for many practical quantum algorithms in various settings~\cite{jordan_Quantum_2012,chowdhury_Quantum_2017,jordan_Quantum_2019,egger_Credit_2021,wilkens_Quantum_2023,li_Potential_2023,skavysh_Quantum_2023,morimoto_Continuous_2023}. 
We utilize a tensor network-based method~\cite{ran_Encoding_2020, holmes_Efficient_2020, garcia-molina_Quantum_2022, zhou_Automatically_2021,gundlapalli_Deterministic_2022, dov_Approximate_2022,iaconis_Quantum_2023,melnikov_Quantum_2023,gonzalez-conde_Efficient_2024,rudolph_Synergy_2023, iaconis_Tensor_2023,jobst_Efficient_2023,jumade_Data_2023,sano_Quantum_2024}, especially the Tree Tensor Network (TTN) ansatz~\cite{shi_Classical_2006}, which is often used to represent quantum states with a 1D structure.
%For a $D$-dimensional multivariate probability distribution discretized into $2^n$ grids per dimension, a state preparation circuit comprising a total of 
%$nD$ qubits is required, which is challenging to compile in general.
We prove that a multivariate normal distribution can be efficiently represented as a TTN
when there is a 1D structure in the correlations between variables.
We demonstrate that by combining the FSL method and Tensor Cross Interpolation (TCI)~\cite{oseledets_TTcross_2010,savostyanov_Quasioptimality_2014},
it is possible to scalably prepare quantum states for multivariate normal distributions with 1D structures, 
independent of the dimension $D$ and discretization parameter $n$.

Additionally, we propose a new approach that combines circuit compilation with automatic structural optimization~\cite{hikihara_Automatic_2023}.
Automatic structural optimization is a method to optimize network structure in TTN to represent the state more efficiently by reconnecting the legs of the local tensor throughout the process.
This allows for the appropriate detection of structures in multivariate normal distributions and finding more compact circuits,
even when the correlations between variables are unknown beforehand or when there is no 1D structure.

We have demonstrated our method to compile state preparation circuits for several high-dimensional multivariate normal distributions. 
Our method can scalably generate state preparation circuits for multivariate normal distributions with 1D structures. 
We also numerically calculate the relationship between the fidelity of the circuit generated and the bond dimension, confirming the theoretical results. 
Furthermore, we investigated the impact of the automatic structural optimization algorithms on the performance of circuit compilation. Our numerical results show that the algorithm appropriately detects the 1D structure of a multivariate normal distribution and reconfigures the network with high success probability, even for large $D$.
%Furthermore, we demonstrate that the automatic structural optimization algorithm can detect the 1D structure of multivariate normal distributions with high probability, even for large $D$.
Finally, we show that we can reduce the circuit depth and CNOT count by factors of $10$ to $1000$ compared to existing methods for random multivariate normal distributions.

This paper is organized as follows. We describe tensor network techniques used in this study in Sec.~\ref{sec: tensor network methods}. We explain the proposed state preparation method in Sec.~\ref{sec: methods}.
In Sec.~\ref{sec: results}, we present the numerical results of our methods and discuss their performance. We conclude the paper by summarizing our contribution in Sec.~\ref{sec: discussion}.

\section{Tensor network methods}\label{sec: tensor network methods}
In this section, we briefly explain the tensor network methods used in this study. 
For a more detailed discussion on tensor networks, please refer to the reference~\cite{bridgeman_Handwaving_2017}.

\subsection{Matrix product states and tree tensor networks}
Tensor networks are a framework for representing a large tensor in the form of a network of smaller tensors. 
Specifically in quantum information theory, they are used to represent the wave function of quantum many-body systems. 
Matrix Product States (MPS), one type of tensor network ansatz, can be described as follows:
\begin{equation}
\ket{\psi_{\mathrm{mps}}} = \sum_{\{s_i\}}\sum_{\{\alpha_i\}}\left[A_{\alpha_1}^{[1]s_1}A_{\alpha_1\alpha_2}^{[2]s_2}\dotsm A_{\alpha_{n-1}}^{[n]s_n}\right]\ket{s_1s_2\dotsm s_n},
\end{equation}
where $\{s_i\}$ and $\{\alpha_i\}$ are referred to as physical and virtual indices, respectively. The dimension of the virtual indices is referred to as {\it bond dimension}.

The ability to represent a multi-dimensional array as an MPS can be evaluated using its singular values. 
Suppose $\psi(s_1,\dotsc,s_i;s_{i+1},\dotsc,s_n)$ be the unfolding matrix between site $(1,\dotsc,i)$ and $(i+1,\dotsc,n)$. 
Consider the singular value decomposition (SVD) of the matrix in the form
\begin{equation}
    \psi(s_1,\dotsc,s_i;s_{i+1},\dotsc,s_n)=U_i\Sigma_iV_i \label{eq:SVD}
\end{equation}
and its singular values $\{\sigma_j^{(i)}\}$ sorted in descending order. 
Note that the summation of the square of the singular values is $1$ for a wave function $\ket\psi$. 
Then, there exists an MPS approximation $\tilde\psi$ with bond dimension $r$ whose truncation error, in the sense of the Frobenius norm,
is upper bounded by its singular values~\cite{oseledets_TTcross_2010,oseledets_TensorTrain_2011}:
\begin{align}
    &\norm{\psi(\bm{s}) -\tilde\psi(\bm{s})}_F \leq \sqrt{\sum_{i=1}^{n-1} \epsilon_i^2}  \label{eq:truncation_error_MPS}, \\ 
    &\epsilon_i^2 = 1 - \sum_{j=0}^{r-1} {\sigma_j^{(i)}}^2.
\end{align}
In other words, a multi-dimensional array is well-represented by an MPS if it has a small number of dominant singular values for each unfolding matrix.

The discussion above can be generalized to Tree Tensor Networks (TTN)~\cite{shi_Classical_2006}.
A TTN is a type of tensor network ansatz that does not have closed loops in the diagrammatic representation. The inner edge divides the sites into two disjoint sets, which we call the {\it bipartition} of the system.
Similarly to the case with MPS, the error upper bound in the best TTN approximation, in terms of the Frobenius norm, 
depends on the distribution of the singular values of the unfolding matrix across each bipartition induced by the edges in the network.

\subsection{Functional matrix product states and tree tensor network}
For the discussion later, we introduce the functional version of MPS, known as functional MPS (functional Tensor-Train decomposition)~\cite{oseledets_Constructive_2013,bigoni_Spectral_2016,gorodetsky_continuous_2018,rohrbach_Rank_2020}.
For $\psi\in L^2(\mathbb{R}^n)$, we consider the following decomposition:
\begin{equation}
    \psi(x_1,\dotsc,x_n)=\sum_{\{\alpha_i\}}\gamma_1(\alpha_0,x_1,\alpha_1)\gamma_2(\alpha_1,x_2,\alpha_2)\dotsm\gamma_n(\alpha_{n-1},x_n,\alpha_n). \label{eq:Schmidt}
\end{equation}
To consider the error introduced by this approximation, we use an analogue of the singular value decomposition: the Schmidt decomposition.
The Schmidt decomposition of the function $\psi$ has the form
\begin{equation}
    \psi(x_1,\dotsc,x_i,x_{i+1},\dotsc,x_n)=\sum_k\sqrt{\lambda_k^{(i)}}\psi_k^{(1)}(x_1,\dotsc,x_i)\psi_k^{(2)}(x_{i+1},\dotsc,x_n), \label{eq:truncation_error_fMPS}
\end{equation}
where $\psi_k^{(1)}$ and $\psi_k^{(2)}$ form orthogonal basis for each subsystem, and $\{\sqrt{\lambda_k^{(i)}}\}$ are called {\it Schmidt coefficients} and sorted in descending order.
Then, as in the discrete version, there exists a functional MPS approximation $\tilde{\psi}$ with bond dimension $r$ whose truncation error, in the sense of 
the $L^2$ norm, is 
upper bounded by its Schmidt coefficients:
\begin{equation}
    \|\psi-\tilde{\psi}\|_{L^2(\mathbb{R}^n)}\leq \sqrt{\sum_{i=1}^{n-1}\left(\sum_{k=r}^\infty\lambda_k^{(i)}\right)} \label{eq:summation}.
\end{equation}
These formulations for the functional MPS can similarly be extended to the functional TTN.

Consider tensorizing a function evaluated at points on a grid. 
When discretized on a sufficiently dense and large grid, the Schmidt decomposition \eqref{eq:Schmidt} can be interpreted as the singular value decomposition \eqref{eq:SVD},
and the truncation error of the functional MPS \eqref{eq:truncation_error_fMPS} as that of discrete MPS \eqref{eq:truncation_error_MPS}.
Therefore, if a function can be accurately approximated by a functional MPS, we can expect that its discretized multi-index tensor can similarly be approximated by a discrete MPS with the same bond dimension. 
In the following discussion, we consider functional MPS and functional TTN when evaluating approximation error.

\subsection{Tensor cross interpolation}
We can construct MPS or TTN from the original multi-index array by sequentially applying SVD. However, this method requires loading the unfolding matrix of the original tensor into memory, which becomes impractical when the tensor to be approximated itself is very large.

Instead, we can use the tensor cross interpolation (TCI) algorithm~\cite{oseledets_TTcross_2010}, a method for directly constructing MPS and TTN from the elements of the multi-dimensional tensor. 
Let $s_i\in\mathbb{K}_i=\{1,\dotsc,d\}$. We define the following multi-indices:
\begin{equation}
    s_{\leq i}=s_1s_2\dotsm s_i\in\mathbb{K}_1\times\mathbb{K}_2\times\dotsb\times\mathbb{K}_i,\quad s_{> i}=s_{i+1}\dotsm s_n\in\mathbb{K}_{i+1}\times\dotsb\times\mathbb{K}_n.
\end{equation}
We interpolate the multi-index tensor as:
\begin{align}
    \psi(s_1,\dotsc,s_n)\simeq &\psi(s_1,\mathcal{J}_{>1})[\psi(\mathcal{J}_{\leq 1},\mathcal{J}_{>1})]^{-1}\psi({\mathcal{J}_{\leq 1},s_2,\mathcal{J}_{>2}})
    [\psi(\mathcal{J}_{\leq 2},\mathcal{J}_{>2})]^{-1}\\ 
    &\dotsm\psi(\mathcal{J}_{\leq n-1},s_n),
\end{align}
where $\mathcal{J}_{\leq i}$ and $\mathcal{J}_{>i}$ are sets of multi-indices satisfying the following condition:
\begin{align}
    \mathcal{J}_{\leq i+1}\subset \mathcal{J}_{\leq i}\times\mathbb{K}_{i+1},\quad \mathcal{J}_{>i}\subset \mathbb{K}_{i+1}\times\mathcal{J}_{>i+1}.
\end{align}
If the size of each $\mathcal{J}_{\leq i}, \mathcal{J}_{>i}$ is limited by $\chi$,
the original tensor is interpolated as an MPS with bond dimension $\chi$. 
The choice of multi-indices defines the accuracy of the approximation.
In this study, we use the ALS maxvol algorithm to find indices to use. For the details of the algorithm, please refer to the reference~\cite{oseledets_TTcross_2010}.

\subsection{Canonical form of tree tensor network}\label{sec:canonical}
\begin{figure}[t]
    \centering
    \includegraphics[width=0.7\linewidth]{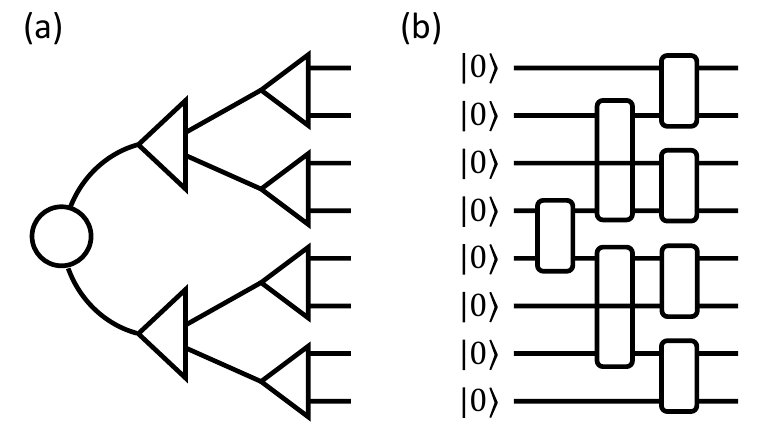}
    \caption{(a) The canonical form of the TTN. The top tensor is illustrated as the circle node and isometries as the triangle nodes. (b) The corresponding quantum circuit.}
    \label{fig:TN}
\end{figure}

For tree tensor networks, an efficient truncation method using the {\it canonical form} is well-known~\cite{schollwoeck_densitymatrix_2011}.
The diagram notation of the canonical form is illustrated in Fig.~\ref{fig:TN}(a). We impose the isometric constraint on each tensor, except for the order-2 tensor known as the {\it top tensor}.
We can efficiently transform any TTN into canonical form by repeatedly applying SVD to each tensor.
When sweeping the position of the top tensor in canonical form, we can truncate small singular values and reduce bond dimension, which achieves a near-optimal approximation in terms of the Frobenius norm.

After obtaining the canonical form for the 1D tensor network states, we can easily construct the state preparation circuit as illustrated in Fig.~\ref{fig:TN}(b). The quantum circuit representation of the top tensor is achieved through the brute-force state preparation method. The other tensors are isometries and can be converted into circuits using unitary synthesis~\cite{iten_Quantum_2016}.
Even if the bond dimension is not a power of two, it can be represented as a quantum circuit by suitably padding tensors to become a power of two.

\section{Methods}\label{sec: methods}
Our goal is to compile a quantum circuit to generate a quantized $D$-dimensional multivariate normal distribution by discretizing each dimension into $2^n$ points.
In this study, we consider the following probability density function with mean zero for simplicity: 
\begin{equation}
    p(\bm{x}) = \frac{1}{\sqrt{(2\pi)^D \det(\Sigma)}} \exp\left(-\frac{1}{2} \bm{x}^\top \Sigma^{-1} \bm{x}\right),
\end{equation}
where $\bm{x}$ is $D$-dimensional vector and $\Sigma$ is the covariance matrix. Then we specify the parameter $a$ such that $[-a/2,a/2]^D$ contains effectively all of the probability density.
We use the binary encoding to discretize each dimension, such that
\begin{equation}
    x_i = -\frac{a}{2} + \frac{a}{2^n}b_i,\quad 0\leq b_i\leq 2^n-1,
\end{equation}
then our target state is expressed as follows:
\begin{equation}
\sum_{\bm{b}}f_{\bm{b}}\ket{\bm{b}}\equiv\sum_{\bm{b}}\left(\frac{a}{2^n}\right)^{D/2}\sqrt{p(\bm{x})}\ket{\bm{b}}.
\end{equation}

\subsection{Fourier series approximation}
To scale for $n$, we adopt the technique of using discrete Fourier transform as proposed in ~\cite{moosa_Lineardepth_2023}. 
The discrete Fourier transform over $\mathbb{Z}_{2^n}^D$ is expressed as follows:
\begin{equation}
    \hat{f}_{\bm{k}}=\frac{1}{\sqrt{2^{nD}}}\sum_{b_1=0}^{2^n-1}\dotsm\sum_{b_D=0}^{2^n-1}f_{\bm{b}}\exp\left(-2\pi i\frac{b_1k_1+\dotsb+b_Dk_D}{2^n}\right).
\end{equation}
Then we approximate $f_{\bm{b}}$ by truncating Fourier coefficients as follows:
\begin{equation}
    \tilde{f_{\bm{b}}}\simeq\frac{1}{\sqrt{2^{nD}}}\sum_{k_1=-M/2}^{M/2-1}\dotsm\sum_{k_D=-M/2}^{M/2-1}\hat{f}_{\bm{k}}\exp\left(2\pi i\frac{b_1k_1+\dotsb+b_Dk_D}{2^n}\right).
\end{equation}
Here, $M:=2^m$ represents the number of Fourier coefficients to be kept for each dimension. It is shown that, if $f(\cdot)$ is a sufficiently smooth function, the required number of $m$ is smaller than $n$ and, moreover, does not depend on $n$~\cite{moosa_Lineardepth_2023}. In practice, $\hat{f}_{\bm{k}}$ is calculated as follows:
\begin{align}
    \hat{f}_{\bm{k}}&\propto \sum_{\bm{b}}f_{\bm{b}}\exp\left(-2\pi i\frac{x_1k_1+\dotsb+x_Dk_D}{a}\right)\exp\left(-2\pi i\frac{k_1+\dotsb+k_D}{2}\right) \\
    &\propto(-1)^{|k_1+\dotsb+k_D|}\int_{-\frac{a}{2}}^{\frac{a}{2}}\dotsi\int_{-\frac{a}{2}}^{\frac{a}{2}}\sqrt{p(\bm{x})}\exp\left(-2\pi i\frac{\bm{xk}}{a}\right)d\bm{x}
\end{align}
for a sufficiently fine discretization. Moreover, if $a$ is taken to be sufficiently large, the following approximation also holds:
\begin{equation}
    \hat{f}_{\bm{k}}\propto(-1)^{|k_1+\dotsb+k_D|}\exp\left(-(2\pi/l)^2\bm{k}^T\Sigma\bm{k}\right) \label{eq:A_k}
\end{equation}
with an appropriate normalization term. Equation \eqref{eq:A_k} shows that the value of $\hat{f}_{\bm{k}}$ decreases exponentially as the wavenumber $\bm{k}$ increases, indicating that retaining
only a very small number of Fourier coefficients is sufficient to represent $f_{\bm{b}}$.

We first prepare the state that encodes the Fourier coefficients:
\begin{equation}
    \sum_{k_1=-M/2}^{M/2-1}\dotsm\sum_{k_D=-M/2}^{M/2-1}\hat{f}_{\bm{k}}\ket{\bm{k}}, \label{eq:fk}
\end{equation}
then we apply Inverse Quantum Fourier Transformation (QFT) to obtain the target state $\sum_{\bm{b}}\tilde{f_{\bm{b}}}\ket{b}$.
The circuit cost for preparing \eqref{eq:fk} is much smaller than that for $\sum_{\bm{b}}f_{\bm{b}}\ket{b}$, and
Inverse QFT can be implemented efficiently in quantum computers with circuit depth $n$.
Therefore, we can generate the desired state with a small cost that does not scale exponentially with the discretization size $n$.

\subsection{Representing Fourier coefficients using tree tensor network}
We now want to encode the truncated Fourier coefficients $\hat{f}_{\bm{k}}$ into the quantum circuit. In the case of $D$-dimensional probability distributions, however, the number of truncated Fourier coefficients is $2^{mD}$, which can be a very large number.

Therefore, we consider representing the Fourier coefficients with an MPS:
\begin{equation}
\sum_{\bm{k}}\hat{f}_{\bm{k}}\ket{\bm{k}} = \sum_{\{k_i\}}\sum_{\{\alpha_i\}}\left[A_{\alpha_1}^{[1]k_1}A_{\alpha_1\alpha_2}^{[2]k_2}\dotsm A_{\alpha_{D-1}}^{[D]k_D}\right]\ket{k_1k_2\dotsm k_D},\label{eq:fk_mps}
\end{equation}
or a TTN. The memory footprint required grows only polynomially with $D$.

\subsubsection{Schmidt coefficients of multivariate normal distribution}
As discussed above, the accuracy achievable in the MPS or TTN representation depends on its singular values or Schmidt coefficients. 
The Schmidt coefficients for a multivariate normal distribution are analytically known~\cite{bogdanov_Schmidt_2016}.

First, we discuss the Schmidt decomposition of the two-dimensional normal distribution. The Schmidt decomposition of the wave function $\psi(x_1,x_2)$ has the form
\begin{equation}
    \psi(x_1,x_2)=\sum_{k=0}^{\infty}\sqrt{\lambda_k}\psi_k^{(1)}(x_1)\psi_k^{(2)}(x_2).
\end{equation}
Suppose the quantum state $\psi(x_1,x_2)$ is given by the square root of the density function of the normal distribution with correlation coefficient $\rho$:
\begin{align}
    \psi(x_1,x_2) &= \sqrt{p(x_1,x_2)}, \\
    p(x_1,x_2) &= \frac{1}{2\pi \sigma_1 \sigma_2 \sqrt{1-\rho^2}} \exp\left(-\frac{1}{2(1-\rho^2)}\left(\frac{(x_1 - m_1)^2}{\sigma_1^2} \right.\right. \nonumber \\
    & \qquad \left.\left. - 2\rho\frac{(x_1 - m_1)(x_2 - m_2)}{\sigma_1 \sigma_2} + \frac{(x_2 - m_2)^2}{\sigma_2^2}\right)\right),
\end{align}
where $m_1,m_2$ are the mean values and $\sigma_1^2,\sigma_2^2$ are the variance of each variable. The Schmidt decomposition of this wave function~\cite{bogdanov_Schmidt_2007} is given by:
\begin{align}
    \psi_k^{(1)}(x_1)=C_k^{(1)}H_k\left(\frac{(x_1-m_1)}{\sigma_1}\sqrt{\frac{K}{2}}\right)\exp\left(-K\frac{(x_1-m_1)^2}{4\sigma_1^2}\right), \\
    \psi_k^{(2)}(x_2)=C_k^{(2)}H_k\left(\frac{(x_2-m_2)}{\sigma_2}\sqrt{\frac{K}{2}}\right)\exp\left(-K\frac{(x_2-m_2)^2}{4\sigma_2^2}\right),
\end{align}
where $C_k^{(1)},C_k^{(2)}$ are the normalization factors and $K$ is calculated by
\begin{equation}
    K=\frac{1}{\sqrt{1-\rho^2}}. \label{eq:K}
\end{equation}
The squares of Schmidt coefficients $\{\lambda_k\}$ form a geometric series with an initial value
\begin{equation}
    \lambda_0=\frac{2}{K+1} \label{eq:lambda_0}
\end{equation}
and common ratio:
\begin{equation}
    q=\frac{K-1}{K+1}. \label{eq:q}
\end{equation}
An approximation error of an MPS of bond dimension $r$
is calculated as 
\begin{align}
    \|\psi - \psi_{\mathrm{mps}}\|_{L^2(\mathbb{R}^2)}^2&\leq \sum_{k=r}^{\infty}\lambda_k \\
    &=\lambda_0\sum_{k=r}^\infty q^k \\
    &=\lambda_0\frac{q^r}{1-q}=q^r.
\end{align}
Then, if we want to suppress the truncation error to $\epsilon$, we need the bond dimension $r$ that satisfies:
\begin{align}
    &q^r \leq \epsilon^2 \\
    \iff &r \geq \frac{2}{\log\frac{1}{q}}\log{\frac{1}{\epsilon}}.
\end{align}
In other words, the required bond dimension scales as $O(\log(1/\epsilon))$, which is an efficient approximation.

The distribution of the Schmidt coefficients of the multivariate normal distribution is also known and calculated only from the covariance matrix~\cite{bogdanov_Schmidt_2016}.
For simplicity, we consider the multivariate normal distribution with zero mean.
The Schmidt decomposition of the multivariate function for subsystems $x_1,\dotsc,x_p$ and $x_{p+1},\dotsc,x_D$, assuming that $2p\leq D$, is defined as follows:
\begin{equation}
    \psi(x_1,\dotsc,x_p,x_{p+1},\dotsc,x_D)=\sum_k\sqrt{\lambda_k}\psi_k^{(1)}(x_1,\dotsc,x_p)\psi_k^{(2)}(x_{p+1},\dotsc,x_D).
\end{equation}
According to the canonical-correlation analysis~\cite{ray_Advanced_1984}, linear transformations can be found
for variables $\xi_1,\dotsc\xi_p,\xi_{p+1},\dotsc,\xi_D$:
\begin{alignat}{2}
    &\xi_i=\sum_{j=1}^{p}l_{ij}x_j,\quad &&1\leq i\leq p \\
    &\xi_i=\sum_{j=p+1}^{D}m_{ij}x_j,\quad&&p+1\leq i\leq D
\end{alignat}
that satisfy the following properties:
(1) All $\xi$ have unit variance and zero mean and 
(2) the correlation between any $\xi$ is zero except for the pairs $(\xi_1,\xi_{p+1}),(\xi_2,\xi_{p+2}),\dotsc,(\xi_p,\xi_{2p})$.
The coefficients $\bm{l},\bm{m}$ and the canonical correlation coefficients $\{\rho_i\}$ between these pairs are the solutions 
to the following generalized eigenvalue problem:
\begin{equation}
\begin{pmatrix}
    O & \Sigma_{12} \\
    \Sigma_{21} & O
\end{pmatrix}
\begin{pmatrix}
    \bm{l} \\
    \bm{m}
\end{pmatrix}
=\rho
\begin{pmatrix}
    \Sigma_{11} & O \\
    O & \Sigma_{22}
\end{pmatrix}
\begin{pmatrix}
    \bm{l} \\
    \bm{m}
\end{pmatrix},\label{eq:gep}
\end{equation}
where $\Sigma_{11},\Sigma_{12},\Sigma_{21},\Sigma_{22}$ are the submatrices of the covariance matrix:
\begin{equation}
    \Sigma = \begin{bmatrix}
    \Sigma_{11} & \Sigma_{12} \\
    \Sigma_{12} & \Sigma_{22} \\
    \end{bmatrix}.
\end{equation}
$D-2p$ solutions are zero and the others arise in pairs as $\pm\rho_1,\pm\rho_2,\dotsc,\pm\rho_p$. 
We choose the positive ones. 
Then, the target function
is decomposed as the tensor product of the probability density function of 
the two-dimensional normal distribution $(\xi_1,\xi_{p+1}),\dotsc,(\xi_p,\xi_{2p})$ with correlation $\rho_1,\dotsc,\rho_p$, and one-dimensional normal distribution $\xi_{2p+1},\dotsc,\xi_D$:
\begin{equation}
    \psi(x_1,\dotsc,x_p,x_{p+1},\dotsc,x_D)=\prod_{i=1}^{p}\left(\sum_{k_i}\sqrt{\lambda_{k_i}^{(i)}}\psi_{k_i}^{(1)}(\xi_i)\psi_{k_i}^{(2)}(\xi_{p+i})\right)\cdot \sqrt{p(\xi_{2p+1})}\dotsm\sqrt{p(\xi_D)},
\end{equation}
where $\{\lambda^{(i)}\}$ is the Schmidt coefficients for the pair $(\xi_i,\xi_{p+i})$.
The Schmidt decomposition of the multivariate
normal distribution is given by defining mode $\bm{k}=(k_1,\dotsc,k_p)$ and the corresponding orthogonal basis as
\begin{align}
    &\psi_{\bm{k}}^{(1)}(x_1,\dotsc,x_p)=\prod_{i=1}^p\psi_{k_i}^{(1)}(\xi_i), \\
    &\psi_{\bm{k}}^{(2)}(x_{p+1},\dotsc,x_D)=\left(\prod_{i=1}^p\psi_{k_i}^{(2)}(\xi_{p+i})\right)\cdot \sqrt{p(\xi_{2p+1})}\dotsm\sqrt{p(\xi_D)},
\end{align}
and the Schmidt coefficients are the product of each Schmidt coefficient of the pairs:
\begin{equation}
    \sqrt{\lambda_{\bm{k}}}=\prod_{i=1}^{p}\sqrt{\lambda_{k_i}^{(i)}} \label{eq:prod}.
\end{equation}
Each $\{\lambda^{(i)}\}$ can be calculated from $\rho_i$ and equation \eqref{eq:K}-\eqref{eq:q}.
Therefore, by solving the generalized eigenvalue problem \eqref{eq:gep}, we can enumerate all the Schmidt coefficients of the multivariate normal distribution 
between any bipartition of the system and calculate the required bond dimensions for the tensor network approximation.
The computational cost of calculating the above values only depends polynomially on the dimension $D$, not on the discretization size $n$.

\subsubsection{Schmidt coefficients of a multivariate normal distribution with 1D covariance structure}
The discussion above can be applied to multivariate normal distributions with any covariance matrix and bipartition. However, if we restrict to the 1D covariance structure,
we can guarantee the existence of efficient MPS or TTN representations of the target state.
Note that a priori rank bounds for approximations in the MPS representation for the multivariate normal distribution with weak correlations are studied in ~\cite{rohrbach_Rank_2020},
This work differs in that we directly use the information of the Schmidt coefficients for proof.

\begin{theorem}
    \label{th:1}
    Let $p(\bm{x})$ be a probability density function of a $D$-dimensional multivariate normal distribution with zero mean and covariance matrix $\Sigma$.
    Assume that, for any bipartition of variables induced by an inner edge of the tree tensor network, $\rank\Sigma_{12}\leq l$ where $l$ is constant and the canonical correlations are less than 1.
    For every $\epsilon$ there exists a TTN approximation $\Tilde{\psi}$ of $\psi(\bm{x})=\sqrt{p(\bm{x})}$ with
    \begin{equation}
        \|\psi(\bm{x}) - \Tilde{\psi}(\bm{x})\|_{L^2\left(\mathbb{R}^D\right)}\leq \epsilon \label{eq:th_truncation}
    \end{equation}
    whose bond dimension $\chi$ is upper bounded by
    \begin{equation}
        %\chi=O\left(\operatorname{polylog}\left(\frac{l\sqrt{D}}{\epsilon}\right)\right)
        \chi\leq \left(C\cdot\log\frac{l\sqrt{D}}{\epsilon}\right)^l
    \end{equation}
    for some constant $C$.
\end{theorem}

This theorem shows that the required bond dimension scales only poly-logarithmically in $1/\epsilon$, which indicates this kind of multivariate normal distribution can be efficiently represented as a TTN.

\begin{proof}
From \eqref{eq:summation}, if the sum of the squares of the Schmidt coefficients for all bipartitions is less than $\epsilon^2/D$, then
an MPS approximation that satisfies \eqref{eq:th_truncation} exists.

Let $q=\max_i q_i$, where $q_i$ is the common ratio of the squared Schmidt coefficients for the pair $(\xi_i,\xi_{p+i})$.
Consider keeping the $r^l$-th largest Schmidt coefficients and truncating all other small values. 
The truncation error is upper-bounded as follows:
\begin{align}
    \sum_{k=r^l}^\infty\lambda_k &\leq 1-(1-q^r)^l \\
    &\leq 1-(1-l\cdot q^r)=l\cdot q^r,
\end{align}
where we use the Weierstrass product inequality and $q^r<1$.
If we want to suppress the error by $\epsilon^2/D$, we have to satisfy the following inequality:
\begin{equation}
    l\cdot q^r \leq \epsilon^2/D.
\end{equation}
By solving this we get
\begin{equation}
    r \geq \frac{2}{\log{\frac{1}{q}}}\log{\frac{l\sqrt{D}}{\epsilon}}.
\end{equation}

\end{proof}
\begin{example}
    The multivariate normal distribution with uniform correlation coefficients. Consider the following covariance matrix:
\begin{equation}
    \Sigma=\begin{pmatrix}
        1 & \rho & \rho & \cdots & \rho \\
        \rho & 1 & \rho & \cdots & \rho \\
        \vdots & \vdots & \vdots & \ddots & \vdots \\
        \rho & \rho & \rho & \cdots & 1
    \end{pmatrix},
\end{equation}
with $-1<\rho<1$. Then any bipartition of the system gives rank-one submatrix $\Sigma_{12}$. Therefore, this probability distribution can be efficiently represented as an MPS
with any index ordering.
\end{example}

\begin{example} \label{ex:2}
    The multivariate normal distribution with 1D correlations. Consider the following covariance matrix:
\begin{equation}
    \Sigma=\begin{pmatrix}
        1 & \rho & \rho^2 & \cdots & \rho^D \\
        \rho & 1 & \rho & \cdots & \rho^{D-1} \\
        \vdots & \vdots & \vdots & \ddots & \vdots \\
        \rho^D & \rho^{D-1} & \rho^{D-2} & \cdots & 1
    \end{pmatrix},
\end{equation}
with $-1<\rho<1$. Then, a bipartition of subsystems $x_1,\dotsc,x_i$ and $x_{i+1},\dotsc,x_D$ gives a rank-one submatrix $\Sigma_{12}$. 
Therefore, this probability distribution can be efficiently represented as an MPS with the same ordering.
\end{example}

\begin{figure}[t]
    \centering
    \includegraphics[width=0.8\linewidth]{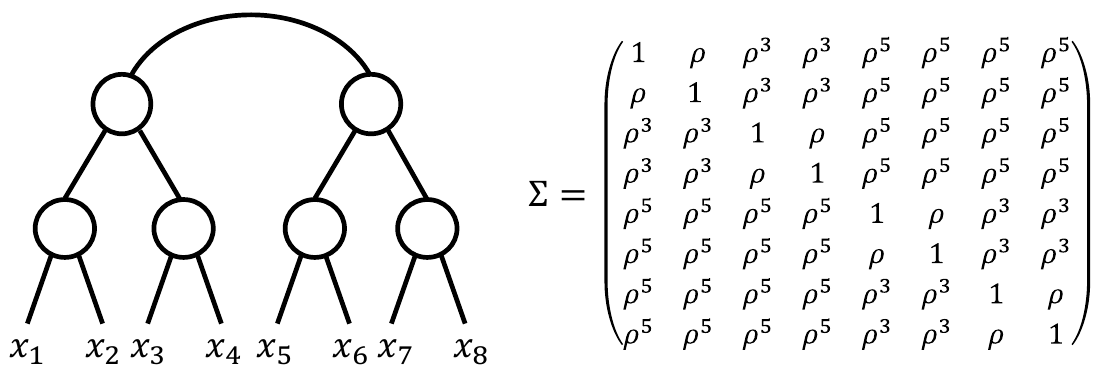}
    \caption{An example of a multivariate normal distribution with a tree-like structure. 
    The covariance decays exponentially with the distance on the tree between variables.
}
    \label{fig:tree-like}
\end{figure}

\begin{example} \label{ex:3}
    The multivariate normal distribution with tree-like correlations. Consider the elements of the covariance matrix are given by its distance:
    \begin{equation}
        \Sigma_{ij}=\exp\left(-\frac{\mathrm{distance}(i,j)}{\sigma}\right),
    \end{equation}
where the distance between site $i$ and $j$ is given as the length of the path on the tree as shown in Fig~\ref{fig:tree-like}. Then, the bipartition of the system induced by each edge of the tree
gives rank-one submatrix $\Sigma_{12}$ and this probability distribution can be efficiently represented as a TTN. 
This generalizes the above two examples.
\end{example}

In addition to the case where the rank of the submatrix is upper-bounded, we can consider the following situation where correlation coefficients decay exponentially.
In this case, however, the required bond dimension for approximation scales polynomially in $1/\epsilon$ instead of $\log(1/\epsilon)$.
\begin{theorem}
    \label{th:2}
    Let $p(\bm{x})$ be a probability density function of a $D$-dimensional multivariate normal distribution with zero mean and covariance matrix $\Sigma$.
    Suppose the canonical correlations are less than 1 and decay exponentially:
    \begin{equation}
        \rho_j\leq \alpha e^{-\theta\cdot j}\quad \alpha,\theta>0
    \end{equation}
    for any bipartition of variables induced by an inner edge of the tree tensor network.
    For every $\epsilon$ there exists a TTN approximation $\Tilde{\psi}$ of $\psi(\bm{x})=\sqrt{p(\bm{x})}$ with
    \begin{equation}
        \|\psi(\bm{x}) - \Tilde{\psi}(\bm{x})\|_{L^2\left(\mathbb{R}^D\right)}\leq \epsilon 
    \end{equation}
    whose bond dimension is upper-bounded by
    \begin{equation}
    %    \chi=O\left(\operatorname{poly}\left(\frac{1}{\epsilon}\right)\right).
        \chi\leq \left(\frac{C_2\sqrt{D}}{\epsilon}\right)^{C_1\log\left(C_3\log\left(\frac{2C_1\log(\frac{C_2\sqrt{D}}{\epsilon})\sqrt{D}}{\epsilon}\right)\right)}
    \end{equation}
    for some constants $C_1,C_2,C_3$.
\end{theorem}

This theorem indicates that the necessary bond dimension scales as $\chi=O(\mathrm{poly}(\frac{C_2\sqrt{D}}{\epsilon}))$ if we ignore the log term in the exponent.
The efficiency of the approximation is worse compared to the case where the rank of the covariance submatrix is strictly upper-bounded.

\begin{proof}
By \eqref{eq:q},
\begin{equation}
    q_j=\frac{K_j-1}{K_j+1}=\frac{\left(1-\sqrt{1-\rho_j^2}\right)^2}{\rho_j^2}<\rho_j\quad(0<\rho_j<1).
\end{equation}
Therefore, when $\rho_j$ decays exponentially, the common ratio $q_j$ also decays exponentially to $j$.

Consider the two-step approximation: First, we retain only the largest singular value for pairs $(\xi_i, \xi_{p+i})$ with $i>l$. Then the approximation error by introducing this is calculated as
\begin{align}
    \|\psi(\bm{x})-\overline{\psi}(\bm{x})\|_{L^2\left(\mathbb{R}^D\right)}^2&=1-(1-q_{l+1})(1-q_{l+2})\dotsm(1-q_p) \\
    &\leq 1-(1-\alpha e^{-\theta(l+1)})(1-\alpha e^{-\theta(l+2)})\dotsm(1-\alpha e^{-\theta p}) \\
    &\leq \sum_{i=l+1}^{p} \alpha e^{-\theta i} \leq \alpha\frac{e^{-\theta l}}{1-e^{-\theta}}.
\end{align}
By solving the following inequality:
\begin{equation}
    \alpha\frac{e^{-\theta l}}{1-e^{-\theta}} \leq \frac{\epsilon^2}{4D},
\end{equation}
we get $l=C_1 \log(C_2\sqrt{D}/\epsilon)$ for constant numbers $C_1,C_2$. 

Next, we consider approximating $\overline{\psi}(\bm{x})$ with finite Schmidt coefficients.
Now the covariance matrix for the multivariate normal distribution $\overline{\psi}(\bm{x})$ only have $l$ non-zero correlation coefficients.
According to theorem \ref{th:1}, we can approximate it with error $\epsilon/(2\sqrt{D})$
by using $r^l$ Schmidt coefficients and $r=C_3\log (2l\sqrt{D}/\epsilon)$ for constant number $C_3$. Then we get
\begin{align}
    \|\psi(\bm{x})-\tilde{\psi}(\bm{x})\|_{L^2\left(\mathbb{R}^D\right)}
    &\leq \|\psi(\bm{x})-\overline{\psi}(\bm{x})\|_{L^2\left(\mathbb{R}^D\right)} + \|\overline{\psi}(\bm{x})-\tilde{\psi}(\bm{x})\|_{L^2\left(\mathbb{R}^D\right)} \\
    &\leq \epsilon/\sqrt{D}
\end{align}

Therefore, the necessary bond dimension to approximate the entire quantum state as TTN within precision $\epsilon$ scales as
\begin{align}
    r^l &=C_3\log\left(\frac{2l\sqrt{D}}{\epsilon}\right)^l \\
    &=C_3\log\left(\frac{2C_1 \log\left(\frac{C_2\sqrt{D}}{\epsilon}\right)\sqrt{D}}{\epsilon}\right)^{C_1 \log\left(\frac{C_2\sqrt{D}}{\epsilon}\right)} \\
    &=\left(\frac{C_2\sqrt{D}}{\epsilon}\right)^{C_1\log\left(C_3\log\left(\frac{2C_1\log(\frac{C_2\sqrt{D}}{\epsilon})\sqrt{D}}{\epsilon}\right)\right)}
%    &=C_3\log\left(\frac{\alpha'\log\left(2\frac{\sqrt{D}}{\epsilon}\right)\sqrt{D}}{\epsilon}\right)^{\alpha'\log\left(\frac{\sqrt{D}}{\epsilon}\right)} \\
%    &=O\left(\left(\log\frac{\log\frac{1}{\epsilon}}{\epsilon}\right)^{\log\frac{1}{\epsilon}}\right).
\end{align}

\end{proof}

\subsection{Automatic structural optimization of the TTN}\label{sec:mps_structure_optimization}
%The distribution of Schmidt coefficients varies depending on the choice of the subsystems; $(x_1,x_2),(x_3,x_4)$ and $(x_1,x_4),(x_2,x_3)$ for example. 
%Therefore, to achieve a more accurate tensor network representation, we need to carefully determine the network structure of the tree tensor network.
As discussed above, the required bond dimension for achieving enough accuracy can be estimated by the distribution of Schmidt coefficients among
variables that are bipartitioned by the network.
Therefore, whether a TTN can efficiently represent a multivariate normal distribution depends strongly on the tensor network structure.
Even if a multivariate normal distribution has a one-dimensional correlation structure as in Th.~\ref{th:1} and~\ref{th:2}, 
an efficient approximation is no longer guaranteed if an appropriate TTN corresponding to that structure is not used.
In the case of a normal distribution, the optimal network structure can be determined by enumerating all network configurations and estimating
their fidelity from the correlation coefficients.
However, for large $D$, it becomes increasingly impractical, as there are $O(D!)$ ways to arrange variables, even when restricted to MPS.
%However, it becomes increasingly impractical for large $D$, even when limiting to MPS, there are $O(D!)$ ways to arrange variables.

\begin{figure}[t]
    \centering
    \includegraphics[width=0.6\linewidth]{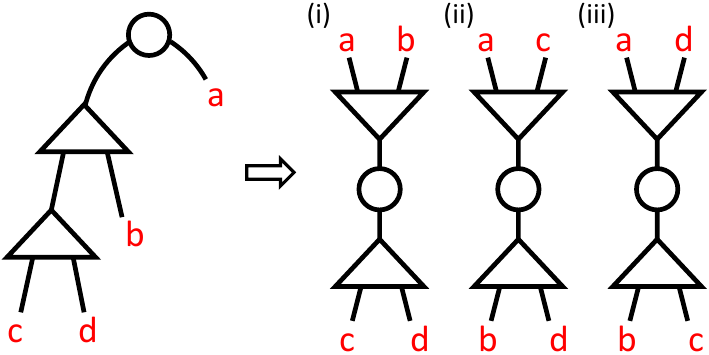}
    \caption{The process of automatic structural optimization of the TTN. When sweeping the top tensor, there are three patterns for choosing indices. By selecting the pattern that results in the lowest entanglement entropy, calculated from the singular values contained in the top tensor, the structure of TTN can be automatically optimized to the most suitable configuration.}
    \label{fig:structual_optimization}
\end{figure}

Instead, in this study, we employ automatic structural optimization~\cite{hikihara_Automatic_2023} to optimize network configuration.
Fig.~\ref{fig:structual_optimization} shows the outline of the algorithm. Initially, we execute TCI using an initial network structure (e.g., MPS) to represent the multivariate normal distribution.
Then, we sweep the top tensor in the network using the canonical form.
During this process, 
%we alter the network structure by choosing one among three bipartition candidates, as shown in Fig.~\ref{fig:structual_optimization}.
%We use entanglement entropy as a criterion for selecting among the candidates.
we contract the three tensors in Fig.~\ref{fig:structual_optimization} and obtain $T_{abcd}$.
We consider three ways of singular value decomposition:
\begin{align}
    T_{abcd}=U_{abe}s_eV_{ecd} \\
    T_{abcd}=U_{ace}s_e'V_{ebd} \\
    T_{abcd}=U_{ade}s_e''V_{ebc}
\end{align}
and compute the entanglement entropy as follows:
\begin{align}
    &EE^{(ab|cd)}=-\sum_es_e^2\log s_e^2 \\
    &EE^{(ac|bd)}=-\sum_es_e'^2\log s_e'^2 \\
    &EE^{(ad|bc)}=-\sum_es_e''^2\log s_e''^2
\end{align}
where we assumed that TTN is normalized.
Among them, we choose the structure with the smallest entanglement entropy. By repeating this process, the entire TTN network is automatically optimized to reflect the entanglement structure of the system.

Entanglement entropy takes small values when the singular values decay rapidly. As discussed previously, the faster the correlation coefficients decay, the more rapidly the Schmidt coefficients decrease. Thus, in this context, entanglement entropy indicates the strength of correlation when dividing variables into two groups. The process of automatic structural optimization can be interpreted as assuming a tree-like correlation structure in the probability distribution
and searching the most suitable network configuration where the correlation between variables is minimized.

We first execute TCI using an initial tree tensor network with a sufficiently large bond dimension enough to accurately represent the multivariate normal distribution. Then, we employ structural optimization to refine the network structure and reduce the bond dimension through truncation using the canonical form. If the accuracy of the state obtained is insufficient, we can rerun TCI using the optimized network structure to achieve higher accuracy.

%This approach is straightforward, 
%but if the dimension $D$ is large, the accuracy of the initial state may be poor, and the accuracy of the 
%final state obtained can also be very low. In that case, we can apply TCI again using the optimized structure.

%In the case of a normal distribution, the optimal network structure can be determined by enumerating all possible index arrangements with respect to the %covariance matrix. However, finding the optimal index ordering becomes challenging for large $D$ or for general multivariate probability distributions.
%Instead, we consider using the technique of automatic structure optimization~\cite{hikihara_Automatic_2023}. In automatic structure optimization, during the truncation process using the canonical form, one chooses from several structures according to the guidelines, as illustrated in Fig.~\ref{fig:structual_optimization}. 
%In this study, we choose the one which has the minimum bipartite entanglement entropy.
%By repeating this process, it is possible to automatically determine a good tree tensor network structure without using information on the strength of correlations between variables directly.

\subsection{Optimizing the quantum circuit by tree tensor network}\label{sec:ttn}
Up to this point, we have discussed how to represent Fourier coefficients using tensor networks efficiently. After encoding the Fourier coefficients as a quantum circuit, 
we can get the discretized probability distribution by applying inverse QFT with a depth of $n$.

%Additionally, it is also possible to optimize the entire state preparation circuit using the canonical form of tensor networks. The circuit for the inverse QFT can be represented as a tree tensor network with bond dimension $2^m$, meaning the entire state preparation circuit
%can be expressed as a tree tensor network. 
%We might get a more compact circuit by optimizing using the canonical form with a smaller bond dimension $\chi<2^m$.

\begin{figure}[t]
    \centering
    \includegraphics[width=0.6\linewidth]{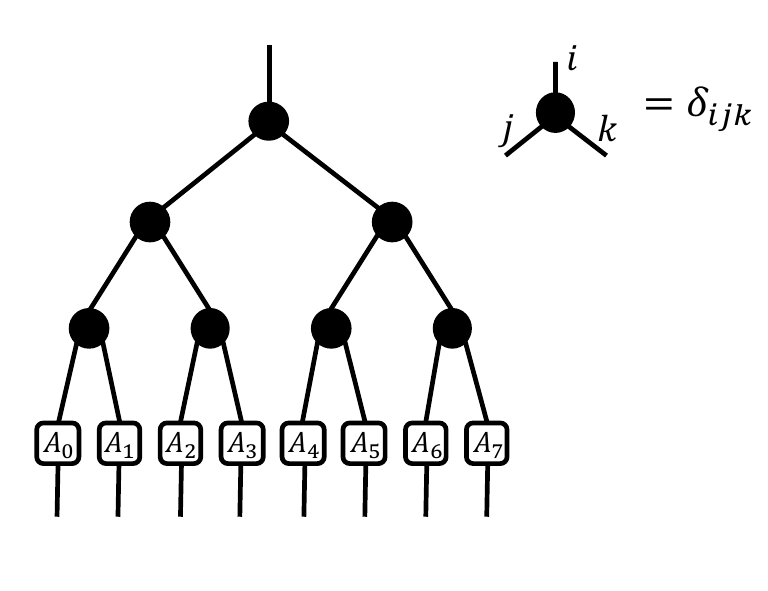}
    \caption{The tree tensor network representing inverse QFT. The black node represents a copy tensor. The definition of tensors $\{A_j\}_{j=0}^n$ is given in the main text. The number of qubits for each dimension is set as $n=8$ in this figure.}
    \label{fig:QFT}
\end{figure}

We can further compile the entire state preparation circuit. 
When expressing Fourier coefficients with an MPS or TTN, a bond dimension $2^m$ is inevitably required for a physical index. This bond dimension can be reduced by utilizing the fact that the inverse QFT circuit for each dimension can be represented as a TTN, as illustrated in Fig~\ref{fig:QFT}. Here, the black node in TTN represents a copy tensor $\delta_{ijk}$ with bond dimension $2^m$ and $A_j$ is a $2^k\times 2$ tensor whose elements are given as
\begin{equation}
    (A_j)_l^k=e^{2\pi ikl/2^{j+1}}.
\end{equation}
We then apply bond truncation algorithms using the canonical form. While this approach loses the advantage of executing inverse QFT with a circuit depth of $n$, it allows for the potential creation of a more compact circuit. Additionally, as the size of each tensor decreases, the classical resources required for the circuit synthesis of each tensor are reduced.

\section{Results}\label{sec: results}
Here we present our simulation results for the compilation method of the state preparation circuit of multivariate normal distribution described in Sec.~\ref{sec: methods}. 
%Our objective is to represent the Fourier coefficients of multivariate normal distributions with various covariance matrix structures using MPS or TTN.
We set the parameters as $a=20$ ($x\in [-a/2,a/2]^D$), $n=8$ (discretizing each dimension into $2^n$ points) and $m=5$ (keep $2^m$ Fourier coefficients for each dimension)
to discretize the probability density function with sufficient accuracy.
We first use TCI to express the Fourier coefficients \eqref{eq:fk_mps} as an MPS with bond dimension 
$\chi'$, where $\chi'$ is set sufficiently large to represent the target Fourier coefficients. In this study, we set 
$\chi'=64$.

%Subsequently, we truncate the bond dimension to $\chi$ using the canonical form described in the previous section. 
The fidelity between the original and approximated states of the $i$-th step of truncation can be calculated using the singular values as follows:
\begin{equation}
    f^{(i)}=\sum_{j=0}^{\chi-1}{\sigma_j^{(i)}}^2.
\end{equation}
Note that we normalize the quantum state after each truncation step so that the summation of the squares of singular values is always one.
After completing the truncation process, the fidelity of the final state is predicted as the product of the fidelity at each step:
\begin{equation}
    f=\prod_i f^{(i)}
\end{equation}
This approach serves as a good estimate. Subsequently, the Frobenius norm is restored using the following equation:
\begin{equation}
    \|\ket\psi - \ket{\tilde\psi}\|_F=\sqrt{2(1 - \sqrt{f})}
\end{equation}

The tensor-network algorithms are implemented using the Python library Quimb~\cite{gray_quimb_2018}. Each data point on the plots is obtained by averaging 100 random instances.
\subsection{Multivariate normal distribution with low-rank structure}
\begin{figure*}[t]
    \centering
    \includegraphics[width=\linewidth]{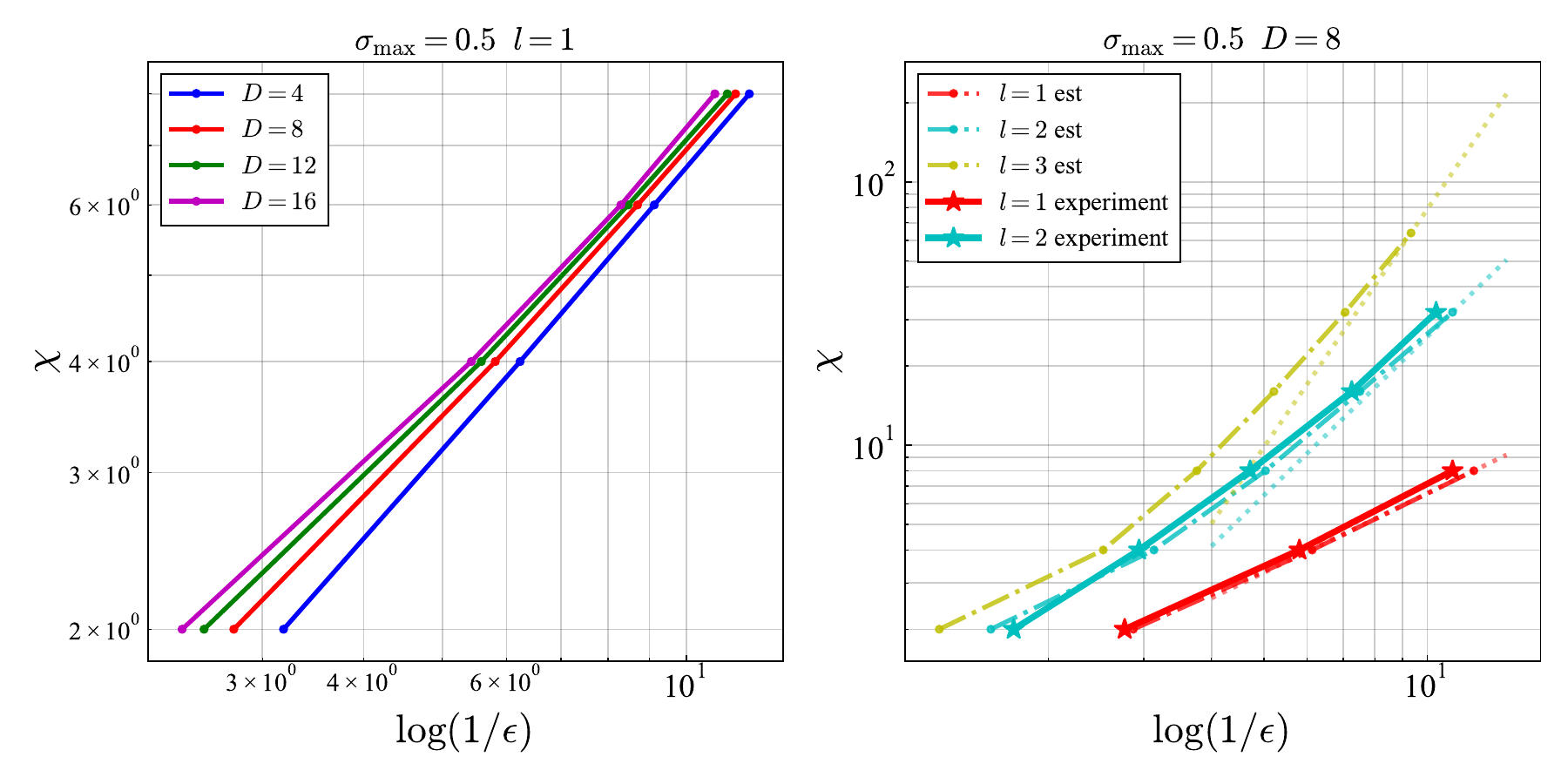}
    \caption{
    The bond dimension of the MPS is plotted as a function of the accuracy $\log(1/\epsilon)$ (Left) for various dimensions of the multivariate normal distribution and 
    (right) for various ranks of the submatrix of the covariance matrix. "est" represents the estimation value from the distribution of the Schmidt coefficients
    and "experiment" represents the actual value from the numerical experiment. The dotted lines represent the predicted polynomial growth rate in $\log(1/\epsilon)$. }
    \label{fig:fig1}
\end{figure*}

To numerically test the theorem \ref{th:1}, we represented a multivariate normal distribution with a low-rank covariance matrix using MPS.
We utilized a simple covariance matrix with a low-rank structure similar to the Example \ref{ex:2}, with
$\sigma$ uniformly sampled from $[0, \sigma_{\mathrm{max}}]$ and $\sigma_{\mathrm{max}}=0.5$.
%We consider a combination of $l$ multivariate normal distributions with a 1D structure.
%We create non-overlapping subgroups of variables $\{x_i^{(j)}\}$ which elements are chosen as $x_i^{(1)}\in\{x_1,x_2,\cdots,x_l\}, x_i^{(2)}\in\{x_{l+1},\cdots,x_{2l}\},\cdots$ for $1\leq i\leq l$. The variables within the subgroup follow
%the multivariate normal distribution with 1D structure. 
We consider a set of $l$ multivariate normal distributions. We partition the variables into non-overlapping subgroups $\{x_i^{(j)}\}$ as follows: for $1\leq j\leq l$, let
\begin{equation}
    x_1^{(j)}\in\{x_1,x_2,\dotsc,x_l\},\quad x_2^{(j)}\in\{x_{l+1},\dotsc,x_{2l}\},\quad\cdots.
\end{equation}
Each subgroup follows a multivariate normal distribution with a 1D structure. For the probability density function constructed in this manner,
the condition on the rank of the covariance matrix $\mathrm{rank}(\Sigma_{12})\leq l$ in Theorem \ref{th:1} is satisfied.

In Fig.~\ref{fig:fig1}, we illustrate the relationship between the accuracy obtained from MPS approximation and the corresponding bond dimension. 
In the left figure, the rank $l$ is fixed at 1 and the dimension $D$ is varied.
As can be seen from the figure, the required bond dimension increases linearly with $\log(1/\epsilon)$. 
Moreover, the required bond dimension grows only gradually as $D$ increases, suggesting that this approach scales effectively with dimension.

In the right plot, the dimension $D$ is fixed, and the rank $l$ is varied. 
The dash-dot line uses the distribution of Schmidt coefficients to directly estimate the achievable accuracy from the bond dimension, 
while the solid line calculates accuracy by obtaining the actual MPS.
The results confirm the Theorem \ref{th:1} that 
the required bond dimension increases with $\log(1/\epsilon)^l$.
Furthermore, the accuracy obtained from the actual MPS is found to be very close to that predicted from the Schmidt coefficients, 
confirming the effectiveness of this bound in estimating the necessary bond dimension given the target accuracy.

\subsection{Multivariate normal distribution with exponentially decaying correlation coefficients}
\begin{figure*}[t]
    \centering
    \includegraphics[width=0.9\linewidth]{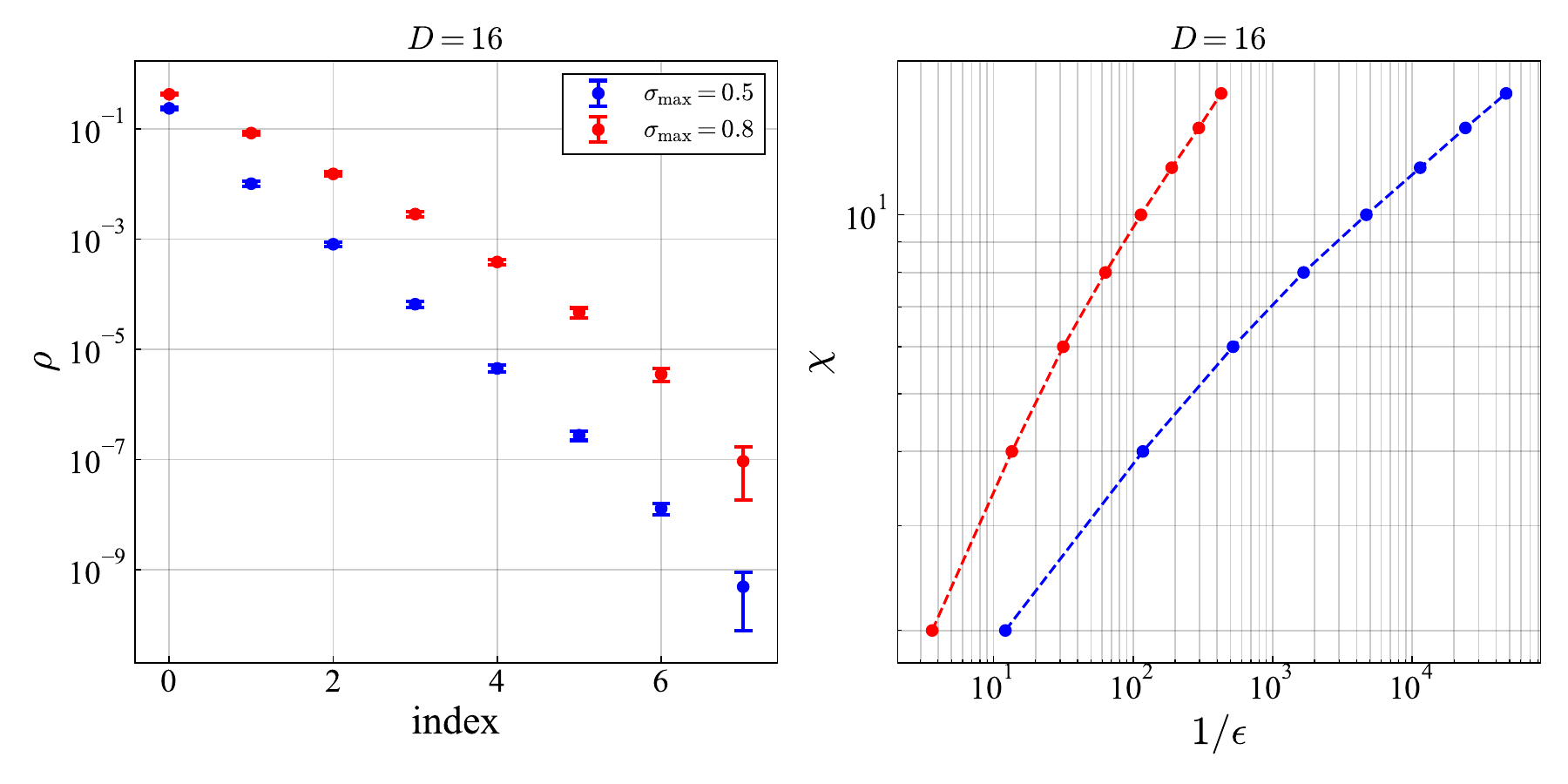}
    \caption{
    (left) The distribution of the canonical correlation coefficients when the system with $D=16$ is divided in half.
    (right) The bond dimension of the MPS is plotted as a function of the accuracy $1/\epsilon$. Different maximum values of covariance $\sigma_{\mathrm{max}}$
    are plotted in different colors.}
    \label{fig:fig2}
\end{figure*}

Next, we represent a multivariate normal distribution in which canonical correlations decay exponentially. To construct such a distribution,
we consider the following setup: the sites are aligned in 1D and the covariance between site $i$ and $j$ is
defined as
\begin{equation}
    \mathrm{cov}_{ij}=(\sqrt{\sigma_i\sigma_j})^{|i-j|},
\end{equation}
where $\sigma_i$ is uniformly sampled from $[0,\sigma_{\mathrm{max}}]$.
We numerically verified that the correlations decay exponentially with the distance between sites. Fig.~\ref{fig:fig2} (left) shows the distribution of canonical correlations between sites $1$ to $8$ and
$9$ to $16$ for the multivariate normal distribution with dimension $D=16$. As can be seen from the figure, the canonical correlations
decay exponentially with the index number, satisfying the condition $\rho_j\leq \alpha e^{-\theta j}$ in Theorem~\ref{th:2}.

Fig.~\ref{fig:fig2} (right) shows the relationship between accuracy and the required bond dimension for approximating the above multivariate normal distribution
with MPS. Note that the $x$-axis represents $1/\epsilon$, not $\log(1/\epsilon)$. As can be seen, polynomial growth in $1/\epsilon$ is confirmed, which is 
consistent with Theorem \ref{th:2}. These numerical results demonstrate that when correlations decay exponentially in 1D,
it is possible to efficiently
represent them using MPS regardless of the dimension $D$, although the efficiency significantly decreases compared to cases when the rank is strictly bounded.

\subsection{Structual optimization with 1D tree structure}
\begin{figure*}[t]
    \centering
    \includegraphics[width=\linewidth]{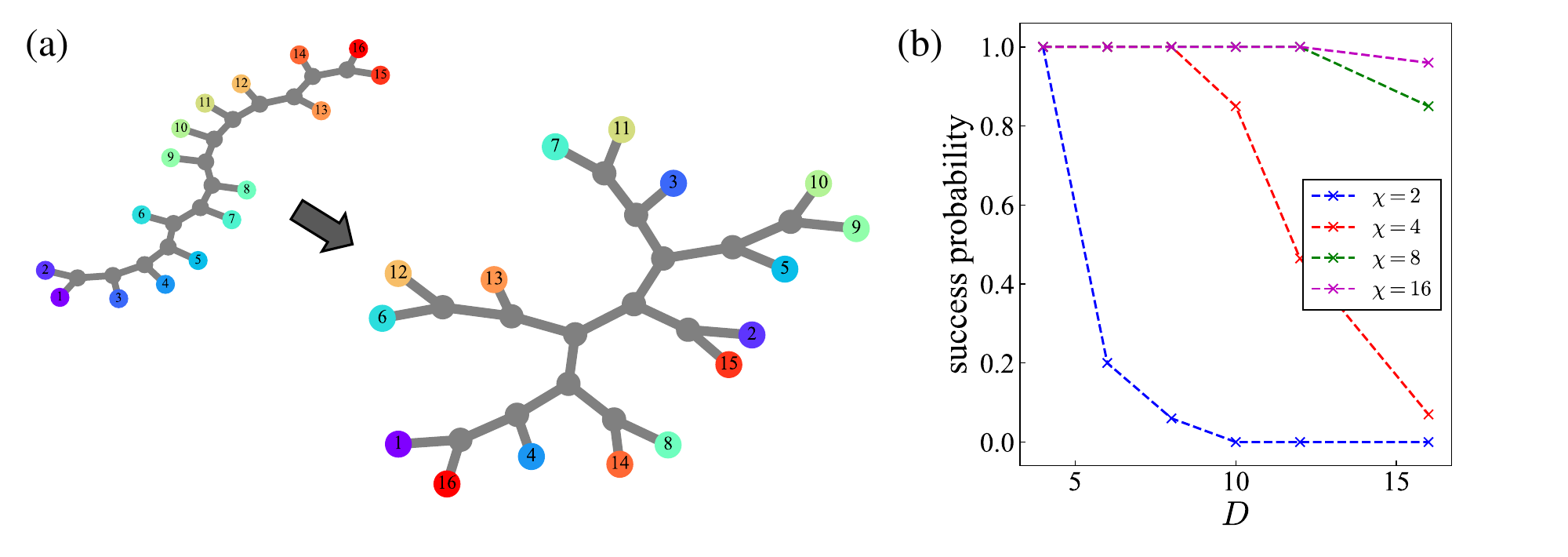}
    \caption{
    (a) Initially, the distribution is represented as an MPS, and after automatic structural optimization,
    the tree-like structure is recovered. (b) The success probability of the structural optimization is plotted as a function of the dimension $D$.
    Different bond dimensions in the tensor network are represented in different colors.}
    \label{fig:fig3}
\end{figure*}

We also evaluated the performance of the structural optimization introduced in Sec.~\ref{sec:mps_structure_optimization}.
We consider a multivariate normal distribution with a tree-like structure as presented in the Example~\ref{ex:3}.
The initial structure of a tree is generated randomly. We approximate the distribution using a standard MPS using TCI without considering the covariance structure.
Then, we employed automatic structural optimization to modify the tensor network, as depicted in Fig.~\ref{fig:fig3} (a), and verify whether it could restore the original tree structure.

Fig.~\ref{fig:fig3} (b) displays the relationship between the dimension $D$ of the multivariate normal distribution and the success rate of structural optimization,
i.e. the probability of completely restoring the original structure. 
Here, $\chi$ represents the maximum bond dimension of the tensor network during the process.
We see that even for high-dimensional distributions like $D=16$,
automatic structural optimization can replicate the original structure with high probability. 
%Additionally, increasing the bond dimension during structural optimization boosts the success rate because it captures the entanglement entropy more accurately.
Additionally, increasing the bond dimension provides more accurate information about the bipartite entanglement entropy of the system, allowing a higher success rate.
These numerical results indicate that despite being a heuristic algorithm, structural optimization is extremely powerful in capturing the 1D structure of distributions.

\subsection{State preparation of random normal distribution}
Now we move on to the compilation of the state preparation circuits for the multidimensional normal distribution without any 1D or tree-like structure.

Before discussing the results, we outline how the generated circuits are evaluated. We assume all-to-all connectivity.
The main metrics adopted here are the number of CNOT gates and the depth of the circuit, both of which are very important, especially for NISQ computers.
Regarding the number of CNOT gates, it is known that an isometry with a p-qubit input and a q-qubit output in the tensor network can be implemented using $O(2^{p+q})$ CNOT gates~\cite{iten_Quantum_2016}. 
For simplicity and comparison purposes, we treat this number as $2^{p+q}$, essentially equating it to the memory size required by the tensor network. 
Circuit depth is the necessary depth for implementing all two-qubit gates and is calculated by identifying the longest path from the root to the leaf within the TTN. If there is an isometry of size $2^{p+q}$ along a path, then $2^{p+q}$ is added to the circuit depth.
Using these metrics, we estimate the cost of the circuits. 
Actual circuit synthesis on quantum hardware is beyond the scope of this work and is left for future investigation.

We consider preparing the quantum state of a 4-dimensional multivariate normal distribution with a randomly generated covariance matrix. Each element of the covariance matrix is generated from the uniform random distribution $[-\sigma_{\mathrm{max}}, \sigma_{\mathrm{max}}]$, except for the diagonal elements, which are all $1.0$ for regularization. 

\begin{figure*}[t]
    \centering
    \includegraphics[width=\linewidth]{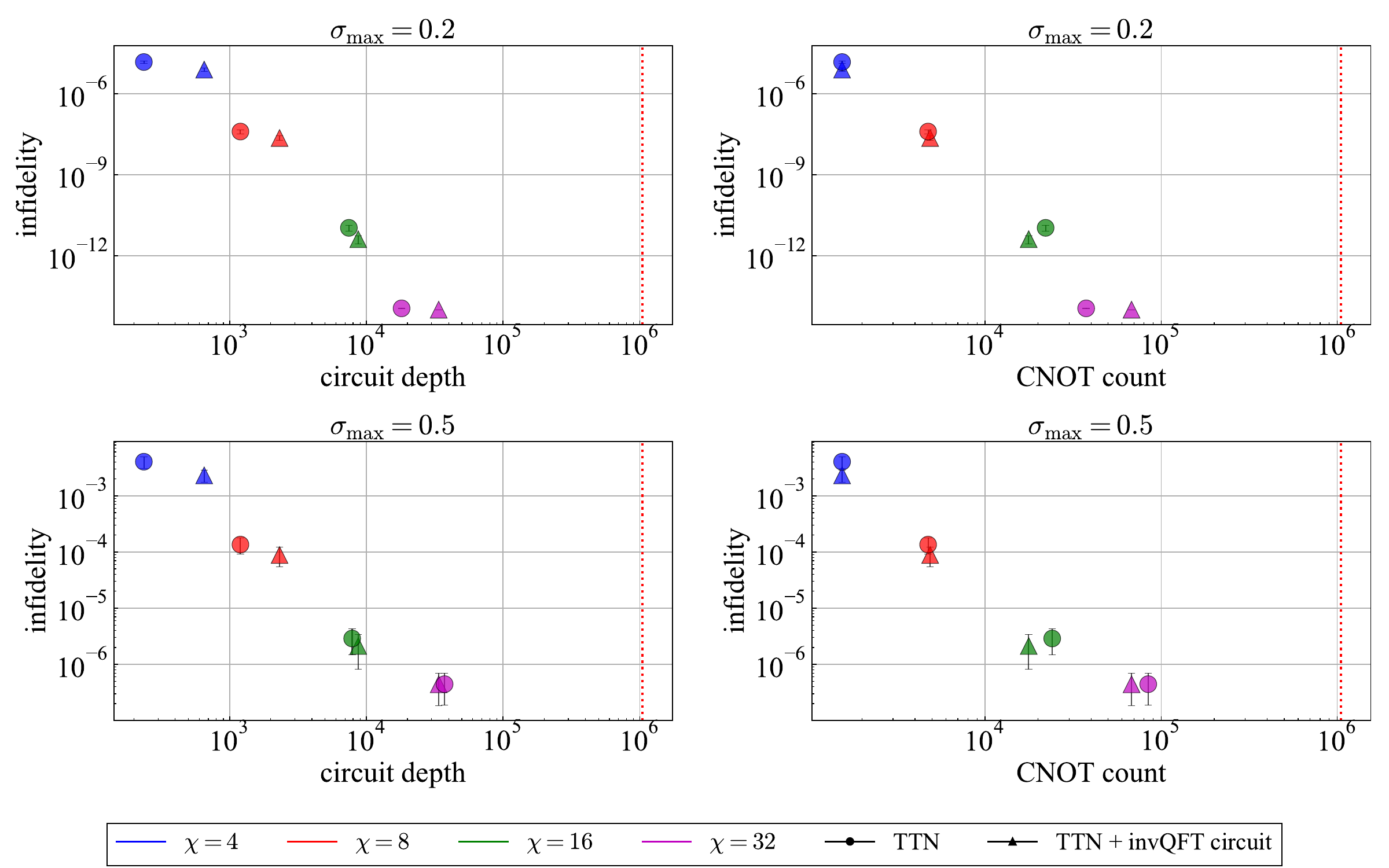}
    \caption{The circuit depth and CNOT count of the encoding circuit, and the infidelity of the quantum state generated by the encoding circuit. 
    The parameters are set as $\sigma_{\mathrm{max}}=0.2,0.5$. 
    The circle points show the result by representing inverse QFT as a TTN, discussed in Sec.~\ref{sec:ttn}, and triangle points show the result by using standard inverse QFT circuit.
    The red line implies the result without MPS approximation.}
    \label{fig:infidelity}
\end{figure*}

In Fig.~\ref{fig:infidelity}, we present the circuit depth, CNOT count, and infidelity of the encoding circuit generated by our method. For comparison, we also plot the costs using the FSL method without tensor network approximations, which serves as a baseline~\cite{moosa_Lineardepth_2023}. As shown in the figure, our approach compiles circuits that achieve high accuracy while using 10 to 1000 times fewer resources than the baseline. Additionally, by using a lower bond dimension, it is possible to perform state preparation at a lower cost, albeit at the expense of accuracy. This is particularly useful for NISQ algorithms, where the required precision is lower, allowing the use of circuits that require very few resources. Notably, at $\sigma_{\mathrm{max}}=0.2$, we can further reduce the circuit depth and CNOT count that optimize the entire circuit through a TTN representation,
as discussed in Sec.~\ref{sec:ttn}.

On the other hand, increasing $\sigma_{\mathrm{max}}$ leads to a significant reduction in accuracy.
$\sigma_{\mathrm{max}}$ represents the strength of the correlations between variables.
Stronger correlations result in larger Schmidt coefficients, making tensor network approximations more challenging.

\begin{figure*}[t]
    \centering
    \includegraphics[width=\linewidth]{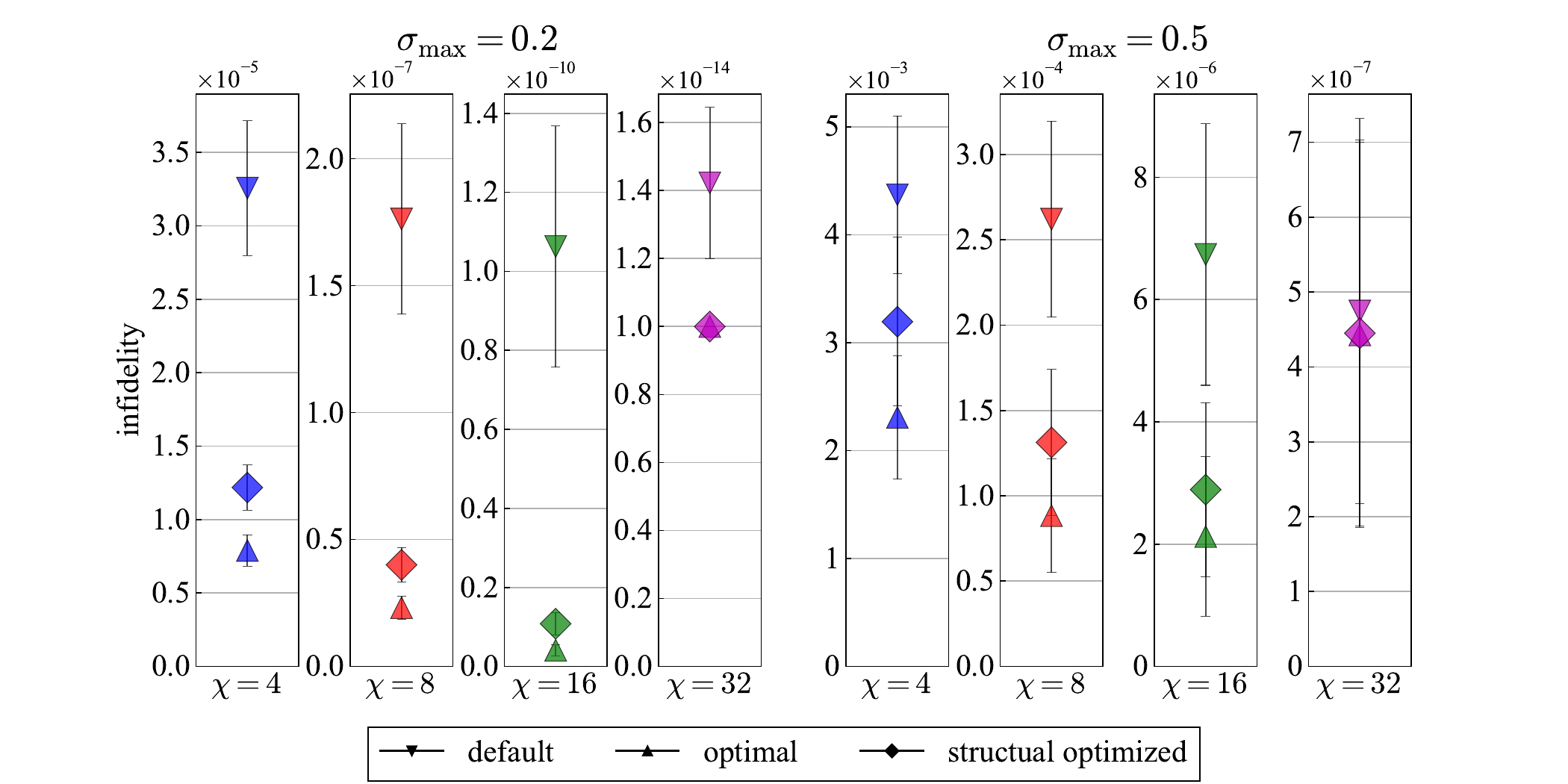}
    \caption{
    The infidelity of the encoding circuit with different strategies for determining network structure.
    The parameters are set as $\sigma_{\mathrm{max}}=0.2,0.5$. 
    The lower triangle points show the result without any optimization of the ordering. The upper triangle points show the result with optimal network structure.
    The diamond points show the result with the automatic structural optimization method discussed in Sec.~\ref{sec:mps_structure_optimization}.}
    \label{fig:ordeing}
\end{figure*}

Finally, we discuss the effect of network structure and automatic structural optimization. 
Fig.~\ref{fig:ordeing} shows the infidelity of the state preparation circuit by varying the strategy for determining network structure.
We get the optimal network structure by enumerating all possible candidates and estimating the fidelity.
It is found that using tensor network structural optimization achieves accuracy close to that of optimal network structure, compared to when optimization is not employed. This highlights the significant influence of network structure on the performance of tensor network-based quantum circuit compilation and 
underscores the potential of structural optimization techniques in finding the near-optimal ordering, even when there are no 1D or tree-like structures.

\section{Discussion}\label{sec: discussion}

In this study, we proposed a scalable circuit compilation method for multivariate normal distributions using TTN. We first considered multivariate normal distributions with 1D correlation structures. We examined two cases: the rank of the covariance matrix is bounded, and the correlation coefficients decay exponentially with the distance between variables. We theoretically guaranteed that these probability distributions can be efficiently represented using TTN.

Additionally, we introduced automatic structural optimization into our circuit compilation method. This optimization rearranges the structure of tensor networks based on the bipartite entanglement entropy of the system. By applying it to multivariate probability distributions, we found that it searches for structures that appropriately reflect the 1D correlation structure of the distribution.

We combined these techniques with the FSL method, TCI, and truncation method using the canonical form. We numerically demonstrated that we can scalably compile the state preparation circuit for normal multivariate distributions with 1D structures. Moreover, even for random normal distributions without a specific structure, we confirmed numerically that our approach could significantly reduce the circuit depth and the number of CNOT gates compared to existing methods.

Multivariate normal distributions play an important role in various practical quantum algorithms. An important example is quantum algorithms for quantum field theory~\cite{jordan_Quantum_2012,jordan_Quantum_2019}. In lattice $\phi^4$ theory, the vacuum state $\ket{\mathrm{vac}}$ is represented as a multivariate normal distribution in the field variables $\phi(\bm{x})$, with the covariance matrix defined as
\begin{equation}
    G(\bm{x-y})=\braket{\mathrm{vac}|\phi(\bm{x})\phi(\bm{y})|\mathrm{vac}}.
\end{equation}
In the free theory, the covariance matrix $G(\bm{x-y})$ decays exponentially with the distance $|\bm{x-y}|$. Therefore, when the lattice is defined in one spatial dimension, the vacuum state can be well approximated using MPS, allowing efficient compilation via our proposed methods.

Another important application lies in quantum algorithms for finance. In finance, the relationship between the risks and returns of multiple assets is modeled with a multivariate normal distribution; therefore, preparing a multivariate normal quantum state is essential for quantum Monte Carlo integration applied to derivative pricing~\cite{sun_Quantum_2024,stamatopoulos_Derivative_2024} and risk analysis~\cite{woerner_Quantum_2018,miyamoto_Quantum_2022,wilkens_Quantum_2023}. These financial datasets are not guaranteed to exhibit exact tree-like correlation structures. However, as demonstrated in our numerical experiments, our heuristic optimization algorithm may help reduce the cost of preparing such quantum states.

One of the advantages of our method is the ability to calculate achievable fidelity in advance. There are a lot of methods for circuit compiling for NISQ algorithms~\cite{ran_Encoding_2020, holmes_Efficient_2020, garcia-molina_Quantum_2022, zhou_Automatically_2021,gundlapalli_Deterministic_2022, dov_Approximate_2022,iaconis_Quantum_2023,melnikov_Quantum_2023,gonzalez-conde_Efficient_2024,rudolph_Synergy_2023, iaconis_Tensor_2023,jobst_Efficient_2023,jumade_Data_2023,sano_Quantum_2024,nakaji_Approximate_2022,zhu_Generative_2022}, which also produce low-cost circuits at the expense of fidelity. However, many algorithms are based on variational optimization and there are often no theoretical guarantees or estimates regarding the accuracy of their approximations. Our method does not depend on variational optimization and can estimate the achievable fidelity from the covariance matrix and bond dimension in the network.
Although this study was limited to multivariate normal distributions with theoretical guarantees, it can be extended to other multivariate probability distributions and multivariable functions~\cite{tindall_Compressing_2024}.

Another advantage of our method is its hardware compatibility. 
While our numerical simulations assume all-to-all connectivity, the hierarchical structure of the TTN allows each local tensor to be encoded into a local quantum circuit and connected using SWAP gates in a layered manner.
If the bond dimension is sufficiently small, the cost of isometry synthesis can also be kept low. 
Therefore, the proposed method is expected to be practical even on quantum hardware with limited qubit connectivity.
We left fully synthesizing circuits and testing them on real hardware for future work.

From the perspective of tensor network algorithms, this research can also be considered a new application of automatic structural optimization. Automatic structural optimization, developed in the field of many-body physics, has been used to detect the entanglement structure of systems. In \cite{harada_Tensor_2024}, the techniques were applied to the field of machine learning, utilizing mutual information to efficiently represent datasets. This work demonstrates that these techniques can also be applied to the field of quantum circuit optimization, underscoring the potential applicability across various fields.

Future directions could involve the use of other tensor network algorithms, especially those designed for two-dimensional systems, 
for the compilation of state preparation circuits. 
Moreover, since automatic structural optimization remains a heuristic algorithm, 
there is a potential to develop theoretically guaranteed or more sophisticated structural optimization algorithms.

An open-source implementation of the method for detecting the correlation structure of multivariate normal distributions
is available and has been described in a subsequent paper~\cite{watanabe_TTNOpt_2025}.

\section*{Acknowledgement}
H. Manabe is supported by JSTCOI-NEXT program Grant Numbers JPMJPF2014. Y. Sano is supported by JSPS KAKENHI Grant Numbers JP23KJ1178. We also thank to the Quantum Computing for Quantum Field Theory summer school 2023 for providing us with the opportunity for collaboration.

\bibliographystyle{plainnat}
\bibliography{manabe_sano}

\onecolumn\newpage
\appendix

\end{document}